\documentclass[preprint,pre,a4paper,superscriptaddress,showpacs,nofootinbib]{revtex4}

\usepackage{graphicx}
\usepackage[body={16cm,22cm},bottom=1.5cm]{geometry}
\usepackage{amsmath}
\usepackage{latexsym}
\usepackage{amsfonts}
\usepackage{psfrag}

\begin{document}
\title{Linear and nonlinear viscoelasticity of a model
unentangled polymer melt: Molecular Dynamics and Rouse Modes analysis}
\author{Mihail Vladkov\footnote{To whom correspondence should be addressed. E-mail: mihail.vladkov@lpmcn.univ-lyon1.fr},
Jean-Louis Barrat}

\affiliation{
Laboratoire de Physique de la Mati\`ere Condens\'ee et Nanostructures,
Batiment L\'eon Brillouin, 43 Bd du 11 Novembre
Universit\'e Claude Bernard Lyon 1 \& CNRS\\
69622 Villeurbanne Cedex, France}

\date{\today}
\setcounter{page}{1}

\begin{abstract}
Using molecular dynamics simulations, we determine the linear and
nonlinear viscoelastic properties of a model polymer melt in the
unentangled regime. Several approaches are compared for the
computation of linear moduli and viscosity, including Green-Kubo and non
equilibrium molecular dynamics (NEMD). An alternative approach,
based on the use of the Rouse modes, is also discussed. This
approach could be used to assess local viscoelastic properties in
inhomogeneous systems. We also focus on the contributions of different
interactions to the viscoelastic moduli and explain the microscopic
mechanisms involved in the mechanical response of the melt to external sollicitation.
\end{abstract}

\maketitle

\newpage
\section{Introduction}
The response of polymer melts to mechanical perturbations,
involving either oscillatory or steady flow, is of great practical
importance, and has been the object of extensive experimental and
theoretical studies.\cite{Ferry,Larson,DoiEdwards} This response
is well known to be viscoelastic, i.e. the storage and loss moduli
exhibit a strong frequency dependence, and nonlinear, with a
typical shear thinning behavior for the viscosity.

On the simulation side, the steady state viscosity and shear
thinning effects have been studied quite extensively in various
configurations for model systems. An extensive review of recent
work is given in reference.\cite{Kroeger_review} A spectacular
success was the obtention of the Rouse-Reptation (or
unentangled-entangled) crossover in the rheological behavior for model
polymer melts of the bead spring type.\cite{Kroeger00} This
crossover is obtained for chain lengths of the order of $N=100$
monomers. According to the popular wisdom in the field, melts with
$N<100$ should therefore be amenable to a description in terms of
the Rouse model, which is considered as a reasonable
phenomenological description of unentangled melts.

Investigation of frequency dependent response are much less
numerous than for steady state viscosity. In fact we are aware of
only one recent study,\cite{Cifre04} with objectives quite
similar to those of the present article. Our aim is to investigate
the mechanical response of model polymer melts submitted to steady
or oscillatory shear, in order to obtain a characterization in
terms of frequency {\it and} amplitude of the solicitation. In
view of the numerical cost of such calculations, our study will be
limited to short chains, but will explore several values of
amplitudes and frequencies, concentrating on relatively low
frequencies.

 As it turns out, a direct assessment of mechanical
properties using non equilibrium molecular dynamics (NEMD)
is very costly from a computational
viewpoint. Hence it is desirable to explore methods that could
provide the same information with a lesser computational effort.
We will in particular explore the possibility of obtaining
viscoelastic properties directly from a study of Rouse modes.
Those, being single chain properties, offer a much better
statistical accuracy than the stress itself, which is a global
property of the system.

The systems under study are briefly described in the next section.
We then discuss the steady state viscosity, both in the linear and
nonlinear regime.
We use three different methods to determine the viscosity - NEMD
simulations, equilibrium Green-Kubo approach and show how the viscosity can be obtained
in a third way from the analysis of equilibrium Rouse modes, provided the
contribution from short times is correctly taken into account.
We then turn to the study of oscillatory strains by means of NEMD
simulations. The conditions
for linear response at low frequency are discussed, and the various
contributions to stress response - storage and loss - are estimated.
Again, the results are compared to an equilibrium analysis based on
the Rouse model.

\section{System description and methods}
The chains are modelled by an abstract and generic, though well
studied, bead spring model - the rather common "Lennard-Jones +
FENE" model.\cite{KremerGrest} All monomers in the system are
interacting through  the Lennard-Jones potential:
\begin{equation}
\label{eq:ljpotcut}
U_{lj}(r) =  \bigg\{ \begin{array}{lll}
         4\varepsilon((\sigma/r)^{12} - (\sigma/r)^{6}), &r\le r_c \\
         0, &r>r_c
       \end{array}
\end{equation}
where $r_c=2.5\sigma$.
Neighbor monomers in the same chain are linked by the FENE (Finite
extension non-linear elastic) potential:
\begin{equation}
\label{eq:FENE}
U_{FENE}(r) =  \frac{k}{2}R_0 \ln(1-(\frac{r}{R_0})^2), \qquad r<R_0
\end{equation}
where $R_0=1.5\sigma$ and $k=30.0 \varepsilon/\sigma^{2}$. The
typical size of the studied systems was of 1280 and 2560 particles
(for chain lengths of 10 and 20) in a periodic cubic simulation
box. The temperature of the melt was fixed in all simulations at
$k_B T = 1$ and the density was $\rho = 0.85$. For the non
equilibrium runs we use the Lees Edwards periodic boundary
conditions coupled with the SLLOD \cite{EvansMorris} equations
of motion. This algorithm allows us to simulate  an unconfined
periodic system under shear. The stress tensor in the melt is
calculated as $\sigma_{\alpha \beta}(t) = - \frac{1}{V} \sum
r_{ij}^{\alpha} F_{ij}^{\beta}$, where $F_{ij}^{\beta}$ the
$\beta$ component of the force between particles $i$ and $j$. The
kinetic contribution of the stress ($\propto \sum
v_{i}^{\alpha}v_{j}^{\beta}$) was also evaluated and found to have
a negligible contribution to the stress as expected at high
density. The simulation time step was taken to be $\delta t = 0.005\tau_{LJ}$,
where $\tau_{LJ}$ is the Lennard-Jones time. All times in the results
are in units of LJ time. The simulations were performed using the LAMMPS code.
\cite{LAMMPS}

\section{Steady state shear viscosity}\label{visco}

 \subsection{NEMD determination}
\begin{figure}[ht]
\centering
\includegraphics[width=8cm]{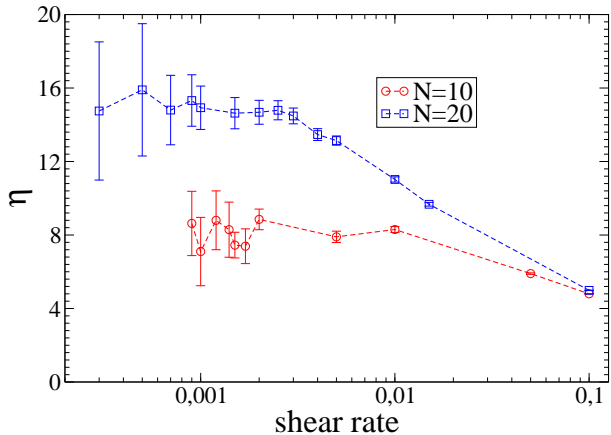}
\caption[Shear viscosity versus shear rate measured by Non
Equilibrium Molecular Dynamics] {Shear viscosity as a function of
shear rate $\dot{\gamma}$ measured in simulations of planar flow
(equation \ref{eqn:eta_sllod}). Shear thinning takes place for
approximately $\dot{\gamma} > 0.01$ for $N=10$ and $\dot{\gamma} >
0.0025$ for $N=20$. The estimated Rouse relaxation times for the
two chain lengths are respectively about $100$ and $400$. Stress
was measured every $2\tau_{LJ}$ for a time ranging from
$15000\tau_{LJ}$ for high shear rates to $100000\tau_{LJ}$ for low
shear rates. Error bars are estimated using block averages for
data correlation.\cite{AT87} \label{fig:eta_sllod}}
\end{figure}

  The most direct  method to obtain shear viscosities is
undoubtedly the nonequilibrium approach that generates
homogeneous, planar Couette flow using Lees-Edwards boundary
conditions and the Sllod algorithm. For a shear rate
$\dot{\gamma}$ and a velocity profile $v_x(y)=\dot{\gamma}y$, the
viscosity is given by
\begin{equation}
\label{eqn:eta_sllod}
\eta = \frac{\langle \sigma (t) \rangle}{\dot{\gamma}}
\end{equation}
Where $\sigma(t)$ is the $xy$ component of the stress tensor in
the sample. The viscosity obtained by this method depends on the
shear rate, generally decreasing with $\dot{\gamma}$ (shear
thinning). An extrapolation is required to estimate the zero shear
rate value. Shear thinning generally takes place when
$\dot{\gamma} > 1/\tau_c$, where $\tau_c$ is a characteristic
relaxation time of the polymer melt (usually, for unentangled melts,
 the Rouse time
$\tau_R$, as can be seen in fig. \ref{fig:eta_sllod}).
\bigskip

 Precise measurements for low shear rates are very time consuming
as substantial statistics are needed for the
 accurate determination of $\langle \sigma (t) \rangle$
 which has a small value for small shear rates. As the
relaxation time of the polymer chains increases rapidly with the
chain length ($\tau_c \propto N^2$), for long chains  very low shear
rates $\dot{\gamma} < 1/\tau_c$, should be used  to reach the
Newtonian  plateau in $\eta(\dot{\gamma})$. This is illustrated by
the large error bars on the low shear rate side of figure
\ref{fig:eta_sllod}. For chains of lengths $N=10$ and $N=20$, our
extrapolation at zero shear rate is consistent with the scaling
expected for unentangled melts, $\eta \propto N$, and  with earlier
estimates found in the literature.\cite{Cifre04}

\subsection{Green-Kubo approach}

The problem of extrapolating the shear rate dependent viscosity
value does not occur in equilibrium methods. The zero shear rate
viscosity is given by the integral of the stress correlation
function (Green Kubo relation)
\begin{eqnarray}
\label{eqn:Gt}
G(t) & = & \frac{V}{k_B T} \langle \sigma_{xy}(t)\sigma_{xy}(0) \rangle \\
\label{eqn:eta_gk}
\eta & = & \int_0^{\infty} G(t)dt
\end{eqnarray}
with $T$ and $V$ the melt temperature and volume. Obtaining an
accurate value for the slowly decaying stress correlation function
involves very long runs in order to have sufficient statistics.
Assuming for simplicity that $\sigma(t)$ obeys Gaussian
statistics, the standard error in the time correlation function at
long times is given by \cite{AT87}
\begin{equation}
\label{eqn:errGK}
\delta(\langle \sigma(t) \sigma(0) \rangle) \approx \sqrt{\frac{2\tau_c}{t_{run}}} \langle \sigma^2 \rangle
\end{equation}
where $\tau_c$ is a characteristic correlation time of the data.
Given that the stress correlation is a rapidly decaying function
(see fig. \ref{fig:str_cors},
 $ \langle \sigma^2 \rangle \approx 10^3\langle \sigma(10) \sigma(0) \rangle$),
this means that if we want to estimate correlations for a correlation time
$\tau_c \approx 100$ it would be necessary to run a simulation of
$10^{8}$ (units of LJ time) in order to obtain a relative precision of $\sim
10\%$ for $\langle \sigma(100) \sigma(0) \rangle$.
The situation in the polymeric case is particularly unfavorable, since the large
viscosity is obtained as the product of a relatively low modulus
and a long relaxation time. Hence the stress correlation function
is long ranged in time, but has a low amplitude, easily masked by
noise associated with the rapid pair-potential part of the stress
that does not involve polymer chain relaxation.
This estimate is indeed very pessimistic and should be seen as an
upper bound of the possible error in the Green Kubo formula.
\cite{Keblinski} In practice the stress autocorrelation
function has several contributions with different weights and
correlation times so that the exact error in the Green Kubo
integral is difficult to evaluate and depends on the chain length.
In our systems of relatively short chains the uncertainty of the
Green-Kubo formula concerning viscosity is of the order of 30\%
when the integral is carried out to several relaxation times and
grows larger as the correlation function is further integrated
(see figure \ref{fig:eta_rouse_int}).

We did not examine the influence of system size on the statistical
accuracy of our data. If a larger system should diminish stress
fluctuations (by a factor $\propto \sqrt{N}$, where $N$ is the
number of particles), a similar effect is expected by longer
simulation time (a factor $\propto \sqrt{T}$). As the
computational effort for larger system increases roughly as
$N\log{N}$, we do not expect to gain better precision for less
computational time this way, so we did not examine the trade off
between system size and run time.

 We are aware of several determinations of polymer
melt viscosity using the Green-Kubo approach. In
\cite{depablo} a good agreement with NEMD results was
obtained for short simulation times without  discussion on
uncertainty. Another  calculation was made in \cite{Sen},
where the authors reported the existence of a large  noise in the
correlation function. This problem was solved by performing
running averages to smooth the data, a method that can reduce the
noise due to rapid bond vibrations but whose effect on the  the
intrinsic statistical accuracy of the stress correlation is not
obvious. Our data still indicate a large error bar (30\%), when
this error bar is  estimated from the three independent components
of the stress tensor.  A viscosity determination based on a
Green-Kubo formula in terms of an Einstein relation was presented
in \cite{Mondello}, unfortunately with relatively little
details that would allow us to compare with our results in terms
of efficiency and accuracy. It is clear that the Green-Kubo
formula remains the only exact way of determining the viscosity
from equilibrium simulations, and should be used whenever an
"exact" result is required. Although this is feasible with a large
computational effort,\cite{Mondello} a detailed report on its
accuracy for polymers is still missing (see however
\cite{Keblinski}). Therefore it seems interesting to discuss
an alternative, faster, approach that can be used for example in
comparative studies at a moderate computational cost.

\subsection{Rouse modes}

It is well known that, at least at a qualitative level, the Rouse
model can account for the viscoelastic behavior of unentangled
polymer melts. Hence, it is tempting to attempt to bypass the
difficulty in obtaining the viscoelastic properties stricto-sensu
by directly using this model. The large uncertainties discussed
above are highly reduced, when we turn to the calculation of
single-particle (or single-chain) correlation functions, which are
the essential ingredient of the Rouse model.
 As the final result
is an average over $M$ separate functions the standard error at
long times should be  $\approx (2\tau_c/Mt_{run})^{1/2}$.

\begin{figure}[ht]
\centering
\includegraphics[width=6cm]{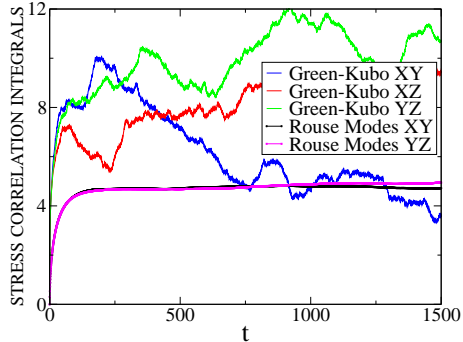}
\caption[Green-Kubo vs Rouse modes stress integrals]
{Integrals of the stress correlation function $G(t)$ for $N=10$
obtained from the global stress (Green-Kubo) and the Rouse modes
stress.  A distinct
plateau that stays stable for very long times ($\sim 50\tau_R$)
is clearly reached with the Rouse modes formulation.
For comparison the Green-Kubo integral value has important
fluctuations for times longer than several times the Rouse
relaxation time $\tau_R$ as predicted by the uncertainty calculation.
Global stress was measured every $0.01\tau_{LJ}$ for a time
span of $25000\tau_{LJ}$ and $G(t)$ was calculated by taking every
point as a starting state and averaging. Such fine sampling is needed to
capture the fast oscillations in $G(t)$ shown in fig. \ref{fig:str_cors}. Rouse
modes were measured every $0.5\tau_{LJ}$ for every chain for a
time of $15000\tau_{LJ}$. Than $G^{Rouse}(t)$ was calculated
averaging over starting states and chains.}
\label{fig:eta_rouse_int}
\end{figure}

The spirit of the Rouse model consists in assuming that the melt
mechanical behavior is dictated by the relaxation of a single
polymer chain, the influence of inter chain interactions being
limited to the phenomenological friction constant. Consistent
with this assumption, the mechanical stress can be calculated
from the Rouse modes of the chains \cite{DoiEdwards}
\begin{equation}
\label{eqn:modestr}
\sigma_{xy}(t) = \frac{\rho k_B T}{N}\sum_{p=1}^{N-1}
\frac{\langle X_{px}(t)X_{py}(t) \rangle}{\langle X_{px}^2 \rangle_{eq}}
\end{equation}
where
\begin{equation}
\label{eqn:rmodes}
X_p(t) = \frac{1}{N}\sum_{n=1}^N r_n (t) \cos{\left( \frac{(n-1/2)p\pi}{N}\right)}\,,\;\;p=0,\ldots,N-1
\end{equation}
are the Rouse modes, with $N$ the chain length, $\rho$ the
monomer number density and $r_n(t)$ - the position of the n-th
monomer in the chain at the time $t$. Assuming, again in the spirit of the Rouse
model, independent Rouse modes, the stress correlation from
equation (\ref{eqn:Gt}) can be rewritten as a function of the
equilibrium correlation functions for individual Rouse  at
equilibrium:
\begin{eqnarray}
G(t) & = & \frac{V}{k_B T} \langle \sigma_{xy}(t)\sigma_{xy}(0) \rangle \\
& = & \frac{V}{k_B T} \frac{1}{T^{sim}} \int_0^{T^{sim}}d\tau \left( \frac{\rho k_B T}{N} \right)^2 \sum_{p,q} \frac{\langle X_{px}(t+\tau)X_{py}(t+\tau) \rangle}{\langle X_{px}^2 \rangle}
\frac{\langle X_{qx}(\tau)X_{qy}(\tau) \rangle}{\langle X_{qx}^2 \rangle} \\
& = & \frac{V}{k_B T} \frac{1}{T^{sim}} \int_0^{T^{sim}}d\tau \left( \frac{\rho k_B T}{N} \right)^2
\frac{1}{N_c^2} \sum_{c,c^{\prime}}
\sum_{p,q} \frac{X_{px}^c(t+\tau)X_{py}^c(t+\tau)}{\langle X_{px}^2 \rangle}
\frac{X_{qx}^{c^{\prime}}(\tau)X_{qy}^{c^{\prime}}(\tau)}{\langle X_{qx}^2 \rangle} \\
& = & \frac{1}{T^{sim}} \int_0^{T^{sim}}d\tau
\frac{\rho k_B T}{N} \frac{1}{N_c} \sum_{c=1}^{N_c}\sum_{p=1}^{N-1}
\frac{X_{px}^c(t+\tau)X_{py}^c(t+\tau) X_{px}^c(\tau) X_{py}^c(\tau)}{\langle X_{px}^2 \rangle} \\
\label{eqn:Grmodes}
& = & \frac{\rho k_B T}{N} \sum_{p=1}^{N-1} \frac{\langle X_{px}(t) X_{py}(t) X_{px}(0) X_{py}(0) \rangle}{\langle X_{px}^2 \rangle^2}
\end{eqnarray}
The viscosity is then calculated using equation
(\ref{eqn:eta_gk}). With this method we observe a substantial
gain in precision (see figure \ref{fig:eta_rouse_int}) and
the values obtained are smaller than the
non equilibrium estimates, by about the same amount for the two
different chain lengths ($\eta_{NEMD}-\eta_{eq} \sim 5$).

\begin{figure}[ht]
\centering
\includegraphics[width=6cm]{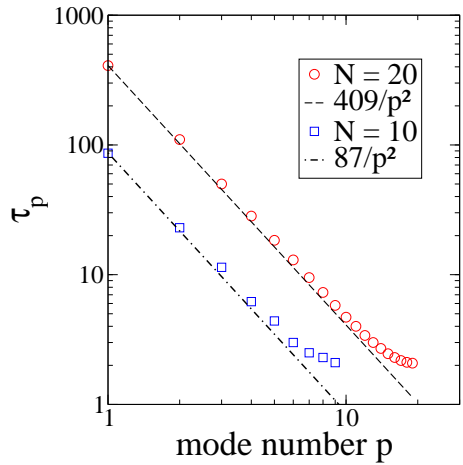}
\caption[Relaxation times of the Rouse modes] {Relaxation times of
the Rouse modes of the chains versus the mode number. Relaxation
times were estimated by an exponential fit of the normalized
correlation function of the mode, which had in all cases an
exponential behavior. Dashed lines show the Rouse theory
prediction $\tau_p \propto 1/p^2$, followed closely by the first
several modes as discussed in \cite{PaulBinder}.}
\label{fig:rouse_times}
\end{figure}

This result is not surprising given the simplifications of the latter
calculation. The interactions between chains in the melt are not
taken into account and the Rouse model cannot yield information
about stress relaxation on very short ``non - polymer'' time
scales.
In order to get a better understanding of the deficiencies in the
calculation using the Rouse model, we first compute the individual
relaxation times of the modes. The relaxation times, shown in
figure \ref{fig:rouse_times}, are extracted from the exponential
decay of these correlation functions. The Rouse scaling $\tau_p
\propto 1/p^2$ is well obeyed for  the first modes.\cite{PaulBinder}
While this is a typical result for the bead-spring model,\cite{PaulBinder}
we note that using more detailed atomistic simulations larger deviations
from the Rouse scaling can be observed, especially in higher modes.
\cite{Harmandaris}
For both chain lengths, the   relaxation time of the fastest mode
was found to be  $\tau_{N-1} \sim 2$.

\begin{figure}[ht]
\centering
\includegraphics[width=7cm]{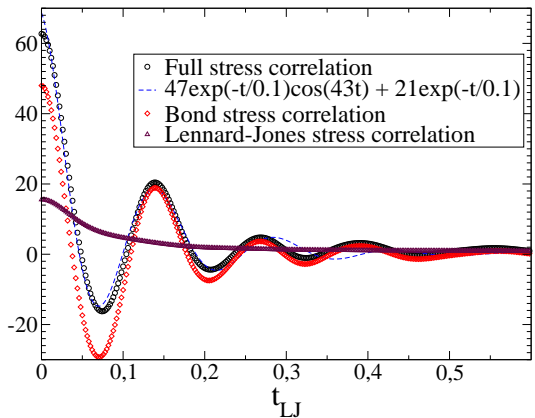}
\caption[Stress correlation function for short times]
{Global stress correlation functions for short times. Correlations of the pair Lennard-Jones and FENE bond
interactions contribution to the global stress are also shown. Dashed line shows a fit of the global stress
correlation to indicate the relevant decay time scales and oscillation frequency.}
\label{fig:str_cors}
\end{figure}

It is clear that the Rouse calculation cannot account for the
contribution to the viscosity associated with time scales shorter
than $\tau_{N-1}$. On such short time scales the Green-Kubo
viscosity integrand can however be obtained with high accuracy, as
illustrated in \ref{fig:str_cors}. This stress-stress correlation
function  first decreases rapidly, and displays a short time
damped  oscillatory behavior,  with a time constant smaller than
 $0.5$. Similar stress oscillations were reported for n-alkanes
as well as for a bead-spring melt.\cite{Mondello,Sen}
In our case, a detailed study of the FENE
  bonds and Lennard-Jones forces
contributions to the stress, shows that  these oscillations are
due  bond vibrations, and that their frequency is close to  the
intrinsic frequency of the FENE bonds.

Integrating this short time stress correlation function gives the contribution to the melt viscosity
of the rapidly decaying part of the global stress, unaccounted for in the Rouse model.
\begin{eqnarray}
\eta & = & \int_0^{\infty} G(t)dt \\
\label{eqn:eta_cor}
& = & \int_{0}^{\infty} G^{Rouse}(t)dt + \int_0^{\tau_{N-1}} \frac{V}{k_B T}
\langle \sigma_{\alpha \beta}(t) \sigma_{\alpha \beta}(0) \rangle dt
\end{eqnarray}

Adding the contribution of the  second term of equation
\ref{eqn:eta_cor} to the viscosity  from the Rouse model provides
a significantly more accurate estimate of the viscosity,
compared to that  calculated from non equilibrium runs (table
\ref{tab:eta}). The short time behavior of the stress correlation
function is the same for  chain lengths $N=10$ and $N=20$, which
is an evidence that this short times contribution is independent
of chain length and represents the missing ``non-polymeric'' part
of the mechanical properties of the melt. It provides a contribution
to viscosity of the order of the viscosity of a simple LJ fluid at the
considered density and temperature.\cite{Kroeger_review}

\begin{table}[ht]
\centering
\begin{tabular}{|l||c|c|}
\hline
Viscosity  & $N=10$ & $N=20$  \\
\hline
Green-Kubo estimate & $8 \pm 4$ & N/A  \\
Rouse modes value  &  $4.88 \pm 0.28$ & $10.2 \pm 1.0$  \\
Short times correction   & $3.38$ &  $3.43$  \\
NEMD extrapolation     & $8.2 \pm 1.3$ &  $14.9 \pm 1.9$  \\
Corrected Rouse value & $8.26 \pm 0.28$ &  $13.6 \pm 1.0$  \\
\hline
\end{tabular}
\vspace*{0.5cm}
\caption[Viscosity measurements]
{Viscosity values from the different methods. Uncertainties in Green-Kubo and Rouse modes $G(t)$
integrals are estimated by evaluating the $\langle \sigma_{xy} (t)\sigma_{xy} (0) \rangle$,
$\langle \sigma_{xz} (t)\sigma_{xz} (0) \rangle$ and $\langle \sigma_{yz} (t)\sigma_{yz} (0) \rangle$ integrals.}
\label{tab:eta}
\end{table}

Being chain length independent, short times stress relaxation
represents a part in viscosity that diminishes with increasing
chain length, about $40\%$ for $N=10$ and $\sim 30\%$ for $N=20$.
Still it cannot be neglected for predicting mechanical properties
of a melt of unentangled chains. It is equally important
 for long chains at large shear rates, when the melt viscosity becomes
comparable to the one of a simple fluid and can explain deviations from the
stress-optic rule.\cite{kroeger97}

The presented method provides a way to rapidly obtain an estimate
of the viscosity as it requires much less computational time than an
accurate Green Kubo measurement. It is still less straightforward and one needs
to monitor $N$ (where $N$ is the chain length) Rouse modes for
every chain to build the correlation function. The method's main
advantage is that, being based on a single chain quantity, it
is local in nature and can be used to assess local
viscoelastic properties in inhomogeneous systems such as confined
melts or melts with filler particles.

\section{Elastic Moduli}
\subsection{Method}

Like the viscosity,  elastic moduli can be obtained using either
equilibrium or nonequilibrium simulation. If one is interested
only in linear response properties, the moduli can be obtained by
Fourier transforming the Green-Kubo integrand $G(t)$ (equation
\ref{eqn:errGK}) in the form
\begin{eqnarray}
\label{eqn:gp1}
 G^{\prime}(\omega) & = & \omega \int_{0}^{\infty}dtG(t)\sin{\omega t} \\
\label{eqn:gpp1}
 G^{\prime\prime}(\omega) & = & \omega \int_{0}^{\infty}dtG(t)\cos{\omega t}
\end{eqnarray}
This equilibrium determination, however, suffers from the same
drawbacks as the Green-Kubo determination of the viscosity, i.e. a
long time, small amplitude tail of $G(t)$ has to be known very
accurately to obtain reasonable results. That is why we did not
show any result concerning elastic moduli issued by a Green-Kubo method. Instead,
the only practical way of using equations \ref{eqn:gp1} and \ref{eqn:gpp1}
is to start from a ''model'' calculation of $G(t)$, in the sense of the Rouse
modeling described in the previous section.

To obtain information beyond the linear regime, the only
possibility is to effectively do NEMD and submit the sample to an oscillatory
strain, using the standard SLLOD \cite{EvansMorris} algorithm. The strain
is  given by
\begin{equation}
\label{eqn:gamdot}
\dot{\gamma_{0}} (t) = \gamma_0\omega\sin{\omega t} \Rightarrow \gamma_0 (t) = - \gamma_0 \cos{\omega t}
\end{equation}
And we define the response of the system by the usual formulae for
the stress $\sigma(t)$
\begin{equation}
\sigma (t)  =  \int_{-\infty}^{t}
G(t-t^{\prime};\gamma_0,\omega)\dot{\gamma_{0}} (t^{\prime})
dt^{\prime}\label{eqn:sigma1}
\end{equation}
A dependence on the shear amplitude and frequency is indicated in
the response function $G$, to recall the possible existence of
nonlinear effects. The frequency dependent moduli are defined from
the Fourier component of $\sigma(t)$ at the imposed frequency
$\omega$
\begin{equation}
\label{eqn:sigma2} \sigma (t) =
\gamma_{0}(G^{\prime}(\omega,\gamma_0)\sin{\omega t} +
G^{\prime\prime}(\omega,\gamma_0)\cos{\omega t}) +
{\mathrm{harmonics \ \  at}\ \  2\omega, 3\omega..}
\end{equation}
The moduli are formally given by the Fourier transforms of the
response function
\begin{eqnarray}
\label{eqn:gp}
 G^{\prime}(\omega;\gamma_0) & = & \omega \int_{0}^{\infty}dtG(t;\gamma_0,\omega)\sin{\omega t} \\
\label{eqn:gpp}
 G^{\prime\prime}(\omega;\gamma_0) & = & \omega \int_{0}^{\infty}dtG(t;\gamma_0,\omega)\cos{\omega t}
\end{eqnarray}
In practice, $G^{\prime}$  (resp. $G^{\prime\prime}$) is extracted from
the time series for the stress by multiplying the signal by
$\cos(\omega t)$ (resp. $\sin(\omega t)$), i.e.
\begin{equation}
G^{\prime}(\omega,\gamma_0)\int_{0}^{T_r}dt\cos^{2}{\omega t}  =
- \frac{1}{\gamma_{0}}\int_{0}^{T_r}dt\sigma(t)\cos{\omega t} +
G^{\prime\prime}(\omega)\int_{0}^{T_r}dt\cos{\omega t}\sin{\omega
t}
\end{equation}
with $T_r$ the length of the simulation run. Thus we obtain the
storage and loss moduli as a function of stress :
\begin{eqnarray}
\label{eqn:Gprime_l} G^{\prime}(\omega,\gamma_0) & = &
\frac{2}{T_r+\frac{\sin}{2\omega T_r}} \left( - \frac{1}{\gamma_0}
\int_0^{T_r}
 dt\sigma(t)\cos{\omega t} + G^{\prime\prime}(\omega;\gamma_0) \frac{\sin^2{\omega T_r}}{2\omega} \right) \\
\label{eqn:Gdprime_l} G^{\prime\prime}(\omega) & = &
\frac{2}{T_r-\frac{\sin{2\omega T_r}}{2\omega}}\left(
\frac{1}{\gamma_{0}}\int_{0}^{T_r}dt\sigma(t)\sin{\omega t} +
G^{\prime}(\omega;\gamma_0)\frac{\sin^2{\omega T_r}}{2\omega}
\right)
\end{eqnarray}
In the above formulae potential harmonic terms were ignored for
simplicity, their contribution being of order $1/T_r$. In the
limit $T_r \gg 1/\omega$ one has simply
\begin{eqnarray}
\label{eqn:Gprime}
G^{\prime}(\omega;\gamma_0) & = & - \frac{2}{T_r\gamma_0} \int_0^T dt\sigma(t)\cos{\omega t} \\
\label{eqn:Gdprime} G^{\prime\prime}(\omega;\gamma_0) & = &
\frac{2}{T_r\gamma_0}\int_{0}^{T}dt\sigma(t)\sin{\omega t}
\end{eqnarray}
In order to elucidate the role of the different interactions for the elastic moduli we
focus on their respective contribution. The stress can very generally
be separated into an intramolecular stress component associated with FENE bonds and intra
chain Lennard-Jones forces, and a intermolecular stress component associated  with
inter-chain Lennard-Jones interactions.
In the following the moduli issued from NEMD simulations will be discussed in terms of these
two separate intra and inter molecular contributions. During a
NEMD run the full stress in the system as well as its inter and intra
molecular components are stored and then elastic moduli are calculated
as described above.

\subsection{Results}
\subsubsection{NEMD Results}
The first question that we investigated is the  extent of the
linear regime, in terms of the strain amplitude. Nonlinear effects
can in principle be detected by a dependence of $G(\omega)$ on
$\gamma_0$, or by the presence of higher harmonics in the stress
signal.

\begin{figure}[ht]
\centering
\includegraphics[width=7cm]{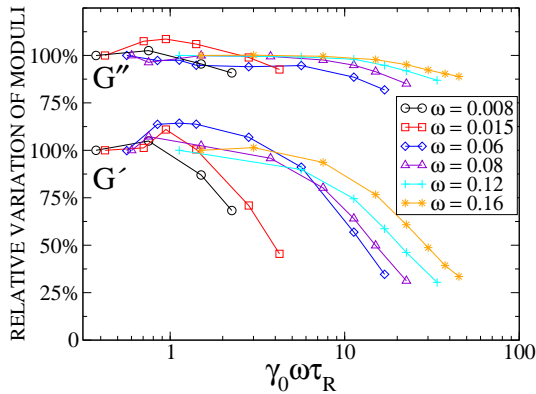}
\caption[Non linear effects in elastic moduli] {Dependence of
elastic moduli on strain rate amplitude $\gamma_0 \omega$ for $N=10$.
Elastic moduli are normalized by their lowest shear rate amplitude value.
For all frequencies the linear regime extends at least to $\gamma_0 \omega \approx 1/\tau_R$,
for $\omega \gg 1/\tau_R$ the linear behavior breaks down for $\gamma_0 \omega \propto \omega$.
The same behavior was measured for $N=20$ (data not shown)}
\label{fig:non_lin}
\end{figure}

There is a clear softening of the response at frequency $\omega$
as amplitude is increased. This softening is obtained above a
value of the strain rate $\gamma_0\omega$ of the order of
$1/\tau_R$ at low frequencies, as illustrated in figure
\ref{fig:non_lin}. The situation is very similar to the shear
thinning behavior of the viscosity, namely the relevant parameter
is the shear rate rather than the strain amplitude or frequency.

 For frequencies $\omega \gg 1/\tau_R$, the softening
 is observed for values of $\gamma_0 \omega$ significantly larger than
 $1/\tau_R$.
  The behavior of chains in this frequency range
is further discussed below.
Harmonics were detected in the time series for $\sigma(t)$ at high
values of the strain amplitude $\gamma_0 > 2.5$ at a frequency of
$3\omega$, where $\omega$ is the solicitation frequency, for all
frequencies. For symmetry reasons the stress must be an odd
function of the strain, so that the response function $G$ is an
even function of $\gamma$.  Hence harmonic contributions are
  observed only for odd multiples of the solicitation
frequency. The amplitudes of the harmonic terms  in the power
spectrum are small, about $8\%$ and $3\%$ of the $\omega$ peak for
respectively $G^{\prime} (\omega)$ and $G^{\prime \prime}
(\omega)$. In the molecular stress harmonics are more visible with
about $17\%$ and $5\%$ for the storage and loss moduli,
respectively. We have not been able to distinguish harmonics for
strain amplitudes $\gamma_0$ below $2.5$. The  observation of
harmonics can thus be attributed to  physical extension of the
chains in which the non linear terms in the interaction potentials
become inevitably important.
Given this preliminary investigation of the non linear regime, we  choose first to
explore the linear response and place ourselves at shear rates
$\gamma_0 \omega < 1/\tau_R$.

\begin{figure}[ht]
\centering
\includegraphics[width=6cm]{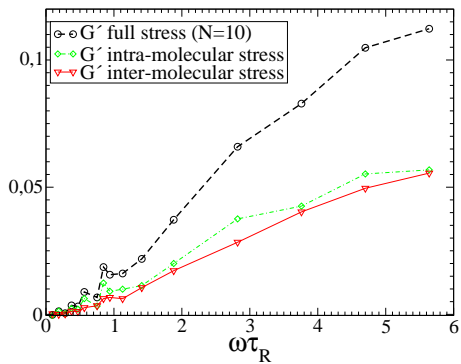}
\caption[Storage modulus as a function of frequency, contribution
of molecular stress] {$G^{\prime} (\omega)$
for $N=10$, measured by NEMD. Contributions from the
intra-molecular and inter-molecular forces are shown. Intra
molecular forces are important for stress storage up to high
frequencies. As $\omega > 1/\tau_p$, the modes $X_i, \; i<p$
are ``frozen'' and behave like stiff springs and store stress
efficiently, so that the intra-molecular component of
$G^{\prime} (\omega)$ grows with frequency.}
\label{fig:G_prime_w_mol}
\end{figure}

\begin{figure}[ht]
\centering
\includegraphics[width=6cm]{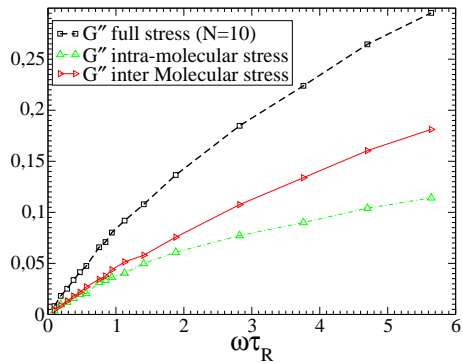}
\caption[Loss modulus as a function of frequency, contribution
of molecular stress] {$G^{\prime \prime}(\omega)$ for $N=10$,
measured by NEMD. Contributions to the
moduli from the intra-molecular and inter-molecular stress
components are also shown. For $\omega > 1/\tau_R$ the
inter-molecular contribution to the loss modulus grows larger than
the molecular component and the system crosses over to a
liquid-like regime.}
\label{fig:G_sec_w_mol}
\end{figure}

Figures \ref{fig:G_prime_w_mol} and  \ref{fig:G_sec_w_mol} display
the frequency dependence of the elastic moduli. As discussed
above, the important uncertainty in $G(t)$ obtained from
equilibrium calculations imposes the use of non equilibrium
methods.

The contributions of the different interactions to the elastic
moduli (calculated by eqn. (\ref{eqn:Gprime}) and
(\ref{eqn:Gdprime})) can be examined by measuring the different
contributions to the global stress. The loss modulus, which is a
measure of stress dissipation in the melt has several
contributions depending on the time scale
(fig. \ref{fig:G_sec_w_mol}):
at short times stress
is relaxed through  the  pair interactions between monomers, in a
"liquid like" manner. At longer times,   relaxation of chain
fragments of length $N_p = 1,2...N$, take place on increasingly
larger time scales. Knowing that a mode $p$ can be viewed as the
relaxation of a subchain of $N_p = N/p$ monomers, for a given mode
$p$ relaxing over $\tau_p$, if the frequency is such that $\omega
> 1/\tau_p$, the mode cannot relax over one oscillation and does
not take part in the stress relaxation, meaning that stress is
relaxed on scales smaller than $N/p$ monomers. This leads to the
Rouse model prediction that, if chain relaxation was the only
process in the melt, the loss modulus had to decrease at high
frequencies. This  decrease was not observed in our model melt
(fig. \ref{fig:G_sec_w_mol}), due to non-polymer relaxation.
As non-polymer relaxation we refer to the relaxation of stress
occuring on a time scale smaller than the relaxation of the fastest
Rouse mode and independent of chain length.
Results show that at high frequencies the behavior of the loss
modulus is dictated by inter molecular interactions, the inter molecular
stress component is dominant. There is a
crossover from polymer melt behavior where stress dissipation is
carried out mainly by chain relaxation to a behavior of a liquid
of interacting chains that do not have time to significantly
change their conformation over one period. In this regime there is
no stress dissipation due to internal polymer chain relaxation and
the increase of the loss modulus with frequency is entirely due to
Lennard Jones pair interactions on very short time scales. The
situation is somewhat different for the storage modulus, where
simulations show that internal polymer chain interactions and
chain modes are important for the elastic response of the melt up
to high frequencies (fig. \ref{fig:G_prime_w_mol}). For a simple
liquid the value of $G^{\prime}(\omega)$ is very small in the
considered frequency range, so it is not surprising that its value
for the melt is due to the chains. The stress storage thus takes
place to a large extent in the slowly varying chain conformations.
Over a large frequency range the slow vibration modes act as an
energy reservoir and the higher the frequency, the more the chains
remain rigid at the time scale of a single period and thus cause
the increase in the storage modulus with frequency. The melt
exhibits an elastic behavior due to the chains that grows stronger
with frequency, in fact the mechanism exposed for the loss modulus
can be applied the other way around for $G^{\prime}(\omega)$. A
given mode $p$ goes rigid as $\omega > 1/\tau_p$ and thus stores
stress (rigid behavior) instead of relaxing it (liquid behavior).
As $\omega > 1/\tau_p$, the modes $X_i, \; i<p$ are ``frozen'' and
take an important part in energy storage. For very high
frequencies $\omega > 1/\tau_{N-1}$ the chain behaves more like a
spring than a flexible polymer.

\begin{figure}[ht]
\centering
\includegraphics[width=7cm]{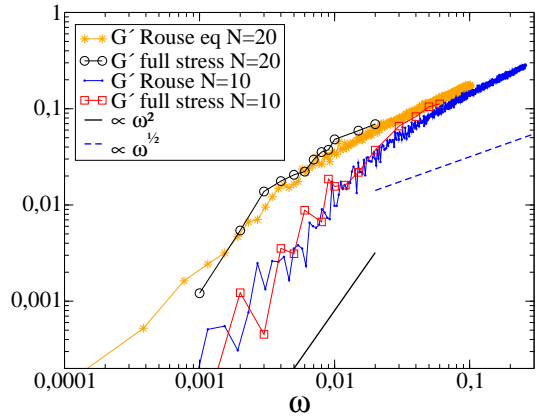}
\caption[Storage modulus from NEMD and equilibrium Rouse calculations]
{Storage modulus from NEMD and equilibrium Rouse modes calculation for chains of
length $N=10$ and $N=20$. For high frequencies the response of both systems: $N=10$ and $N=20$,
becomes identical.}
\label{fig:storage}
\end{figure}

\begin{figure}[ht]
\centering
\includegraphics[width=7cm]{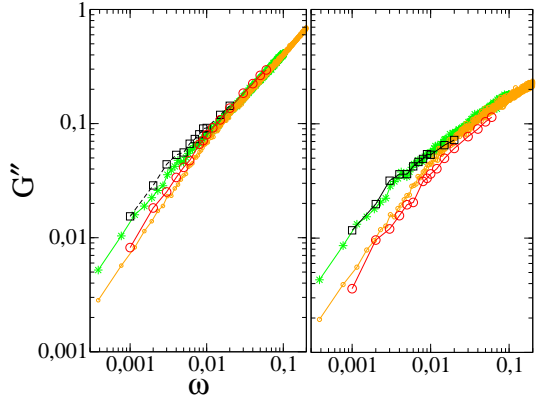}
\caption[Loss modulus from NEMD and equilibrium results from Rouse modes]
{Elastic loss modulus from NEMD full stress (left, $N=20$ - squares, $N=10$ - circles), NEMD intra molecular
stress (right, $N=20$ - squares, $N=10$ - circles), corrected equilibrium Rouse modes calculation (eqn. \ref{eqn:Gpp_w_cor})
(left, $N=20$ - stars, $N=10$ - small circles) and Rouse calculation without short times corrections
(right, $N=20$ - stars, $N=10$ - small circles). The uncorrected Rouse calculation fits well the intra molecular moduli
component.
Stress relaxation for high frequencies is identical for the
two systems, dictated by short subchain and inter molecular forces relaxation on short
time scales as discussed.}
\label{fig:loss}
\end{figure}

\subsubsection{Calculation from Rouse modes and comparison to NEMD data}
\begin{figure}[ht]
\centering
\includegraphics[width=7cm]{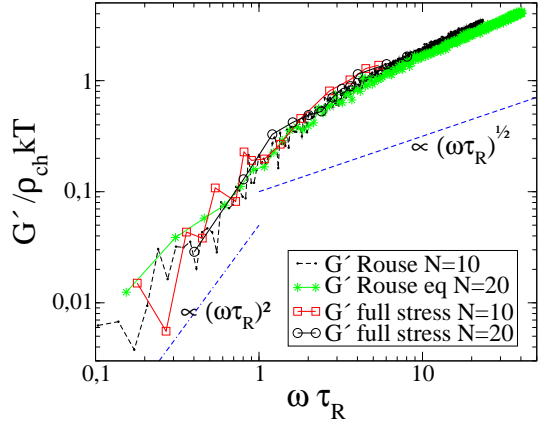}
\caption[Storage modulus scaled by the chain density as a function of reduced frequency]
{Storage modulus divided by the chain density times $k_B T$ as a function of the
reduced frequency $\omega \tau_R$ for the two systems. A cross over in the behavior is
visible for $\omega = 1/\tau_R$.}
\label{fig:storage_scale}
\end{figure}

\begin{figure}[ht]
\centering
\includegraphics[width=7cm]{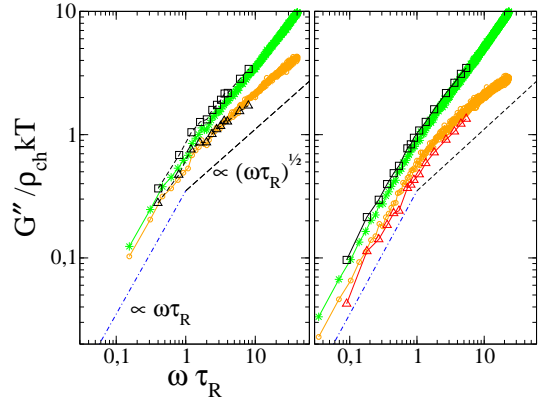}
\caption[Loss modulus scaled by the chain density as a function of
reduced frequency] {Loss modulus divided by the chain density
times $k_B T$ as a function of the reduced frequency $\omega
\tau_R$ for the two systems ($N=20$ - left, $N=10$ - right).
NEMD full stress $G^{\prime \prime}$ - squares, NEMD intra molecular stress
$G^{\prime \prime}$ - triangles, corrected Rouse calculation - stars,
Rouse calculation without short times correction - small circles.
A crossover in the intra molecular component can be seen for $\omega = 1/\tau_R$.
``Non polymer'' relaxation mask the slope change for the full stress
$G^{\prime \prime} (\omega)$.} \label{fig:loss_scale}
\end{figure}

An analytical calculation of $G^{\prime} (\omega)$ and $G^{\prime
\prime} (\omega)$ can be done using the Rouse model
\cite{DoiEdwards,Ferry}. For the bead-spring polymer melt studied
here these functions can be estimated using $G^{Rouse}(t)$ (eqn.
\ref{eqn:Grmodes}) via the integrals in equations (\ref{eqn:gp})
and (\ref{eqn:gpp}). Given that, as discussed above, the stress
storage is dictated by slow, ``frozen'' chain vibration modes for
the whole frequency range, the melt elastic response is well
reproduced by this calculation (see fig. \ref{fig:storage}). As
shown in fig. \ref{fig:loss}, the Rouse approach leads to
an underestimate of the loss modulus, especially at high
frequencies ($\omega > 1/\tau_R$). The increase  in $G^{\prime
\prime} (\omega)$ for high frequencies is due to short time non-polymeric stress
relaxation discussed in the previous section. As discussed in the
first part concerning viscosity, this contribution cannot be
predicted by $G^{Rouse}(t)$ that takes
into account only chain vibration modes. As no inter molecular forces
whatsoever can be taken into account by the single chain Rouse model,
$G^{Rouse}(t)$ provides a good approximation of the loss modulus
calculated from the intra molecular stress component and shows a
discrepancy with the full stress loss modulus that grows with
frequency. In order to obtain an equilibrium single chain
estimate of the mechanical behavior of the melt for all
frequencies, we estimate the short time corrections to
$G^{Rouse}(t)$ in a manner similar to the one used for viscosity.
We use the fit of the short times stress correlation function
(fig. \ref{fig:str_cors}) to calculate the short times correction
to the elastic moduli in the frequency domain by equations
(\ref{eqn:gp}) and (\ref{eqn:gpp}). Writing the short time stress
correlation in the form
\begin{equation}
G^{fast}(t) = A e^{-t/\tau_1} \cos{\Omega t} + Be^{-t/\tau_2}
\end{equation}
leads to
\begin{eqnarray}
\label{eqn:Gp_w_cor}
G^{\prime}(\omega) & = & \omega \int_0^{\infty} G^{Rouse}(t) \sin{\omega t} dt \\
    & + & \frac{A}{2}\left( \frac{\omega(\omega+\Omega)\tau_1^2}{1+(\omega+\Omega)^2\tau_1^2} +
\frac{\omega(\omega-\Omega)\tau_1^2}{1+(\omega-\Omega)^2\tau_1^2} \right) + B\frac{\omega^2 \tau_2^2}{1+\omega^2\tau_2^2} \\
G^{\prime \prime}(\omega) & = & \omega \int_0^{\infty} G^{Rouse}(t) \cos{\omega t} dt \\
&+& \frac{\omega \tau_1 A}{2}\left( \frac{1}{1+(\omega+\Omega)^2\tau_1^2} +
\frac{1}{1+(\omega-\Omega)^2\tau_1^2} \right) + B\frac{\omega \tau_2}{1+\omega^2\tau_2^2}
\label{eqn:Gpp_w_cor}
\end{eqnarray}
where we determine the parameters $A$, $B$, $\Omega$,  $\tau_1$
and $\tau_2$ from the stress correlation function ($A=45$, $\Omega
= 43$, $\tau_1 = \tau_2 = 0.1$ and $B=21$ for both $N=10$ and
$N=20$). Adding these terms to the equilibrium Rouse modes loss
modulus gives a much better estimate of $G^{\prime \prime}
(\omega)$, producing the curves referred to as corrected Rouse (fig. \ref{fig:loss}, corrected
Rouse). The correction concerning $G^{\prime} (\omega)$ is
negligible for the frequency range studied here and the corrected
curve falls on top of the original Rouse curve. This is the
expected result knowing that, as already mentioned, the liquid
like interactions that dominate at short times participate in
stress storage only at very high frequencies.
We find the expected linear
dependence of $G^{\prime \prime} (\omega)$ for $\omega <
1/\tau_R$ (fig. \ref{fig:loss_scale}), the slope being within error bars the value of the
viscosity estimated by planar Couette flow simulations. The
storage modulus has, as expected, $\sim \omega^2$ behavior at
low frequencies ($\omega < 1/\tau_R$) (fig. \ref{fig:storage_scale}).
The Rouse theory predicts a cross over towards a $\propto \sqrt{\omega}$ behavior for higher
frequencies.\cite{DoiEdwards} Our simulations show that $G^{\prime} (\omega)$, estimated by NEMD \textit{and}
Rouse modes measurements, grows slightly faster than $\sqrt{\omega}$ at high frequencies
(fig. \ref{fig:storage_scale}). We can relate this to the discrepancy between theoretical and
measured modes relaxation times and argue that the ``mean field'' presence of multiple chains in
our vibration modes determination promotes more efficient stress storage in the melt at high frequencies.
Simulations show that $G^{\prime \prime} (\omega)$ does not follow the $\sqrt{\omega}$ behavior at
high frequencies either. The loss modulus calculated from the intra molecular stress component,
as well as the direct Rouse
determination follow closely the square root behavior, but the contribution of ``non polymer'' short time
scale inter chain forces change this behavior to almost completely mask the ``polymeric'' cross over.

Following the discussion of the linear properties of the melt we can
further explore the non linear behavior studied by NEMD,
considering the results presented in fig. \ref{fig:non_lin}. Both
the relative decrease in moduli and the harmonics intensity show
that the storage modulus has a stronger non linear behavior
compared to the loss modulus at a given shear rate. As we have
shown that $G^{\prime} (\omega)$ has, at all considered
frequencies, a large intra molecular contribution, and given that
the bond potential is much  steeper than the pair potential acting
between all monomers, it is reasonable to expect that non linear
effects due to large deformations would be more pronounced and
would appear earlier for the storage modulus. From fig.
\ref{fig:non_lin} one can see that for $\omega
> 1/\tau_R$, the higher the frequency, the higher the shear rate
determining the onset of non-linear behavior. This critical value
of the shear rate was found to vary linearly with frequency
$(\gamma_0 \omega)_c \propto \omega$. The system starts behaving
as viscoelastic, and the linear regime extends to higher shear
rates. In fact, as global chain relaxation does not take place on
the time scale of the oscillations, there are already ``frozen''
slow vibration modes in the linear regime at low shear rates. Thus
the relevant shear rate for the onset of non- linearities is
shifted from $1/\tau_R = 1/\tau_1$ to $1/\tau_p$, where $p>1$ is
the number of a higher vibration mode that still relaxes on the
oscillations time scale at the given frequency.

\begin{figure}[ht]
\centering
\psfrag{tr}{$\frac{1}{\tau_R}$}
\psfrag{gw}{$\gamma_0 \omega$}
\psfrag{w}{$\omega$}
\includegraphics[width=7cm, height=6cm]{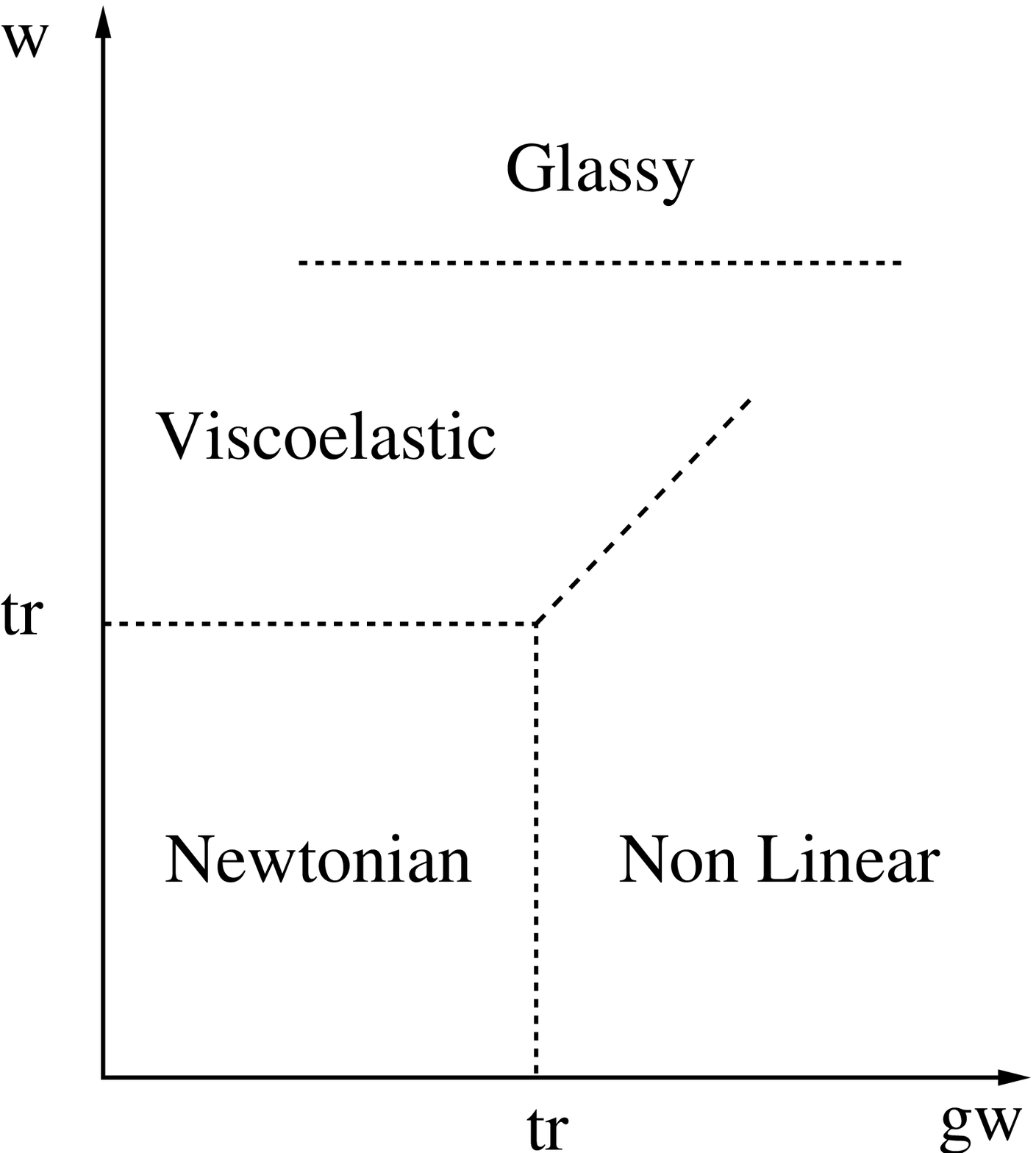}
\caption[Mechanical behavior of the melt for different frequencies and shear rates]
{Mechanical behavior of the melt for different frequencies and shear rates}
\label{fig:summary}
\end{figure}

Finally, we can summarize the overall mechanical behavior of the
studied polymer melt exhibiting several distinct regimes as shown
in fig. \ref{fig:summary}. At low frequencies $0 \le \omega < 1/\tau_R$
and low shear rate amplitudes $\gamma_0 \omega < 1/\tau_R$ the
melt has Newtonian behavior. The viscosity is independent of shear rate
and the elastic response is rather small as most of the stress
is relaxed by the chain conformations. At low frequencies and high
shear rates $\gamma_0 \omega > 1/\tau_R$ the system is non linear: shear
thinning in viscosity, softening in moduli and harmonics in the
measured stress. When we shift to high frequencies $\omega > 1/\tau_R$ the
melt exhibits more pronounced viscoelastic behavior with increasing
storage modulus due to ``frozen'' modes and less intra molecular stress
relaxation. The non linear boundary becomes frequency dependent and is
shifted to higher shear rates determined by the relaxation time scale
of chain segments of length $<N$. The melt is then expected to exhibit
glassy behavior at very high frequencies when no subchain relaxation
whatsoever can occur within an oscillation.

\section{Discussion and Conclusions}

We have discussed the visco-elastic response  of a model
unentangled polymer melt to external shear strain. We use non
equilibrium molecular dynamics methods to directly measure the
elastic moduli and the viscosity of the melt. Our aim was also to
investigate an equilibrium based method for these quantities,
inspired by Green-Kubo relations and possibly offering greater
precision. The  method we proposed is inspired by the Rouse model,
and based on a measurement of the vibration modes of the chains at
equilibrium. This single chain quantity can be rather accurately
measured from an equilibrium simulation. Following the philosophy
of the Rouse model, we use the Rouse modes of the chains to
estimate the long times mechanical behavior of the melt and the
equilibrium stress correlation function to estimate small times,
independent of chain length non-polymeric behavior. The resulting
model provides a satisfactory description of the viscosity and of
the elastic moduli of the melt. The part of the different
contributions in the mechanical stress played in the visco-elastic
melt behavior was also discussed. The NEMD results show that inter
chain interactions in the melt add a non negligible background
part to the storage modulus and dictate the loss modulus behavior
at high frequencies. These interactions should be taken into
account for a precise description of the mechanical properties of
a polymer melt made of relatively short chains. More generally,
these contributions are important for longer chains in several far
from equilibrium situations including relaxation. We measured the
chain Rouse modes mean square equilibrium values and mode
correlations directly from our simulation, so that these
quantities already contain, in a ``mean field'' way, some
information about the background environment of the polymer
chains. After completing this measurements quantitatively with
short times estimate of the stress behavior, the obtained values
for the mechanical properties of the melt are accurate, compared
to non equilibrium ``direct'' measurements. The NEMD results show
that the storage modulus depends on mode relaxation over a large
frequency range, the energy storage takes place in chain
conformations whereas for the loss modulus vibration modes are
``frozen'' one by one as the frequency grows higher than the
inverse chain relaxation time and dissipation is dictated by short
time scales ``non polymer'' stress relaxation. A given vibration
mode $p$ relaxing over $\tau_p$, interpreted as the relaxation of
a sub-chain of length $N/p$ monomers, can contribute mostly to the
loss modulus ($\omega < 1/\tau_p$) or to the storage modulus, if
$\omega
> 1/\tau_p$. This picture provides an explanation for the overall
behavior of the elastic moduli in the studied frequency range. Our
results are in quantitative agreement with previous studies,
\cite{Cifre04} but here we focused on lower frequencies for the
elastic moduli, based on the chain relaxation time that we
estimated, in order to interpret the microscopic mechanisms
involved. Different stress contributions on different time scales
were measured from the simulations, thus allowing the
determination of mechanical response from equilibrium properties
involving Rouse modes measurements and short times corrections.

 We
examined the onset of non linear effects in the measured
quantities, manifested by shear thinning, moduli softening and
harmonics in the stress time series. The non linear regime is
dictated by the shear rate of the solicitation and takes place at
$\gamma_0 \omega > 1/\tau_R$ for $\omega < 1/\tau_R$. At higher
frequencies, onset of nonlinear effects is related to the strain
amplitude rather than rate, as full chain relaxation does not take
place on the time scale of the oscillations. Our study allows a
comprehensive description of the melt mechanical behavior in the
form of a schematic  frequency - shear rate diagram  shown in fig.
\ref{fig:summary}.

We finally mention that the general method presented here is not,
in principle, limited to unentangled melts. Indeed, in the general
reptation picture, the formula used for the stress tensor is the
same as that used in the Rouse model, formula \ref{eqn:Grmodes}.
The relaxation of the modes will be dramatically slowed down by
entanglement effects, so that the viscosity will increase rapidly.
Prefactors, however, are associated with equilibrium correlation
functions and are not affected by entanglement effects.

\bibliographystyle{unsrt}

\end{document}